\documentclass[journal=jacsat,manuscript=article]{achemso}

\usepackage[toc,page]{appendix}
\usepackage[version=3]{mhchem} 
\usepackage{graphicx}
\usepackage{multicol}
\usepackage{setspace}
\usepackage{xcolor}
\usepackage{comment}
\usepackage{amsmath}
\usepackage{nameref}
\usepackage{blindtext}
\usepackage{amssymb}
\usepackage{graphicx} 
\usepackage{caption, subcaption}
\usepackage{upgreek}
\usepackage{tabularx}
\usepackage{multirow} 
\usepackage{xr}
\usepackage{datetime}
\usepackage{hyperref}
\usepackage{ulem}

\def\an#1{\textcolor{gray}{#1}} 

\makeatletter
\newcommand*{\addFileDependency}[1]{
  \typeout{(#1)}
  \@addtofilelist{#1}
  \IfFileExists{#1}{}{\typeout{No file #1.}}
}
\makeatother

\SectionNumbersOn  

\author{Chenghao Zhang}
\affiliation[PNNL]
{Physical and Computational Sciences Directorate, Pacific Northwest National Laboratory, Richland, WA 99352, USA}
\email{chenghao.zhang@pnnl.gov}

\author{Amke Nimmrich}
\affiliation[UW]
{Department of Chemistry, University of Washington, Seattle, WA 98195, USA}

\author{Axel Gomez}
\affiliation[Princeton]
{Department of Chemistry, Princeton University, Princeton, NJ, 08544, USA}

\author{Munira Khalil}
\affiliation[UW]
{Department of Chemistry, University of Washington, Seattle, WA 98195, USA}

\author{Niranjan Govind}
\affiliation[PNNL]{Physical and Computational Sciences Directorate, Pacific Northwest National Laboratory, Richland, WA 99352, USA}
\alsoaffiliation[UW]{Department of Chemistry, University of Washington, Seattle, WA 98195, USA}
\email{niri.govind@pnnl.gov}

\title{A Machine-Learning Framework for Efficient Ring-Polymer Instanton Rate Calculations}

\date{\today}

\begin{document}

\begin{abstract}


We develop an efficient machine-learning framework for ring-polymer instanton rate calculations that combines Gaussian process regression (GPR)-enhanced line integral string optimization with scalable surrogate modeling of the fluctuation prefactor.
By exploiting uncertainty estimates from the surrogate modeling, we show that the number of force evaluations required to converge an instanton path becomes effectively independent of the number of beads used to discretize the pathway. To improve the efficiency of GPR model training, we introduce a strategy combining a physics-informed kernel prior, Hessian-free hyperparameter optimization, and GPU-accelerated Blackbox Matrix-Matrix Multiplication (BBMM), reducing model-training costs by more than an order of magnitude.
For rate calculations, we develop adaptive regression, selective Hessian training, and cubic-spline interpolation strategies that substantially reduce the number of explicit Hessian evaluations while maintaining accurate tunneling rates. We apply and compare both cubic spline interpolation and GPR methods to approximate the instanton rate for representative proton transfer systems such as malonaldehyde, Z-3-aminopropenal, and 7,9-dinitro-10-hydroxybenzo[h]quinoline (dinitro-HBQ). Both approaches perform well for the smaller systems, whereas the dinitro-HBQ results expose limitations of the GPR model and demonstrate the greater robustness of the cubic spline interpolation method. These developments provide a practical workflow for reducing the computational cost of instanton rate calculations in complex molecular systems.

\end{abstract}

\maketitle

\section{Introduction}

Intramolecular proton transfer is a fundamental chemical process that involves coupled electronic and nuclear motion. Including nuclear quantum effects, such as quantum tunneling, is crucial for an accurate description of proton transfer through a barrier. The importance of NQE even at room temperature has been indicated via strong kinetic isotope effect. \cite{Markland2018, Markland2014, HammesSchiffer2005} Direct solution of the Schr\"{o}dinger equation can provide the exact tunneling rate. However, its computational cost scales exponentially with the number of degrees of freedom (DOF), which makes this approach impractical for large molecules. Advanced wave function based methods, including the Multi-Configuration Time-Dependent Hartree (MCTDH) method \cite{MCTDH1990, MCTDHreview, MLMCTDH} and tensor-train methods \cite{soley2021,lyu2022}, have shown promise as practical alternatives owing to their more favorable computational scaling.

Instanton theory \cite{Chapman1975, Miller1975, Coleman1977, Wolynes1981, Weiss1984} provides an efficient way to study tunneling rates in complex molecular systems with sufficient accuracy. In Feynman's path integral formulation, the instanton represents the path that gives the dominant contribution to the tunneling rate. In instanton theory, the tunneling rate is obtained from a path integral that is approximated by the instanton and the harmonic fluctuations surrounding it. Using the isomorphism between equilibrium quantum statistical mechanics and the classical statistical mechanics of a ring polymer, \cite{IsomorphismQM, IsomorphismQM2} the ring polymer instanton theory is formulated by discretizing the instanton path using a ring polymer.\cite{Richardson2009, kastner2014review, Richardson_2018_review} This approach achieves computational efficiency by avoiding explicit sampling of the large number of molecular geometries required in wave function based treatments and enables the inclusion of tunneling effects in atomistic simulations where classical transition state theory fails.

Since its introduction, instanton theory has undergone extensive theoretical developments. A derivation from first principles has been proposed \cite{QTST1, Althorpe2011, Richardson2016}. Extensions to non-adiabatic systems have been developed \cite{WolynesInstanton1, WolynesInstanton2, Cao1997, nonadiabaticInstanton1, nonadiabaticInstanton2, NonadiabaticInstanton3, NonadiabaticInstanton4, NonadiabaticInstanton5}, and several studies have introduced corrections to address rate calculation errors near the cross-over temperature \cite{Weiss1, Cao1996, Kryvohuz1, Kryvohuz2, McConnell2017, PollakCrossOver, PollakCrossOver2, Lawrence2024, Lawrence2025}. Perturbative corrections up to fourth derivatives of the potential have also been incorporated \cite{Lawrence2023, kaser2024}. In addition, microcanonical rate theory \cite{microcanonicalinstanton1, microcanonicalinstanton2, Fang2021} has been formulated within the instanton framework.

On the algorithmic side, the application of ring polymer instanton theory requires locating the instanton path in a high-dimensional quantum system \cite{Rommel2011, Litman2019}, followed by evaluation of the fluctuation factor through Hessians of the potential at the ring polymer beads. Although instanton theory is more efficient than full-dimensional wave function methods, it remains more computationally demanding than classical transition state theory, especially when combined with on-the-fly electronic structure calculations. This computational cost has limited its broader application to complex molecular systems.

Gaussian Process Regression (GPR) based methods \cite{GPR_Instanton, Laude2020, Fang2024, Fang2025} and neural network approaches \cite{NNinstanton, NNInstanton2, Kaser2022} have been developed to accelerate instanton calculations. A chain-of-states approach called the Line Integral Nudged Elastic Band (LI-NEB) method \cite{einarsdottir2012path, LI-NEB2018}, in the same spirit as the NEB method \cite{NEB2000_1, NEB2000_2}, has been introduced for instanton path optimization. In our previous work \cite{LINEB+GPR}, we combined LI-NEB with GPR to locate the instanton path and achieved an order of magnitude reduction in the number of force evaluations required. The LI-NEB and string methods have also been implemented for zero-temperature instanton path searches and tunneling splitting calculations \cite{Cvitas2016, CvitasStringInstanton2018}.

Here, we develop an efficient computational framework for ring-polymer instanton rate calculations that addresses both instanton path optimization and fluctuation prefactor evaluation. Building on our previous GPR-enhanced LI-NEB approach\cite{LINEB+GPR}, we formulate a GPR-enhanced Line Integral String (LI-String) method for efficient instanton path optimization.
The string method \cite{String1, String2} is used to minimize the abbreviated action to obtain the instanton path\cite{Cvitas2016, CvitasStringInstanton2018}. We further show that, when using the surrogate model generated by GPR, the cost of converging the instanton path no longer scales with the number of beads representing the path, consistent with the similar computational scaling reduction that has also been observed for GPR accelerated NEB method \cite{LowScalingGPR2019}. Hyperparameter optimization in GPR based on Cholesky decomposition is inefficient, which limits the applicability of GPR to larger systems \cite{GPR-geo-opt2020, GPR-geo-opt2021, Fang2024}.To improve the efficiency of GPR model training, we demonstrate that an efficient training procedure, a physics-informed kernel prior,
Hessian-free GPR training and GPU-enabled Blackbox Matrix Matrix Multiplication (BBMM) approach \cite{gardner2018gpytorch} can accelerate GPR model training. For intramolecular proton transfer reactions, we find the selective Hessian training strategy \cite{Fang2024} further reduces the cost of Hessian evaluations required for computing ring polymer instanton rates. We also find that the cubic spline interpolation yields a computationally inexpensive yet highly accurate approximation to the instanton rate, which is particularly attractive for 
reducing Hessian costs in higher-dimensional applications.

The paper is organized as follows. In Section~\ref{sec:theory}, for completeness, we briefly introduce ring polymer instanton theory, the LI-String method, and GPR-based surrogate modeling used for path optimization and rate calculations. We also introduce the improved GPR training strategy and emphasize the importance of data quality and outlier data detection. Section~\ref{sec:algoaccel} focuses on algorithmic acceleration of instanton calculations, including low-scaling path optimization, and adaptive regression strategies with selective Hessian modeling. In Section~\ref{sec:Applications}, we apply the proposed approach to representative intramolecular proton transfer systems and finally, we summarize the main conclusions and discuss future directions in Section~\ref{sec:conclusions}.

\section{Theory}
\label{sec:theory}
\subsection{Ring Polymer Instanton Theory}

Semiclassical instanton theory provides a practical way to approximate tunneling rates using the optimal tunneling path, which removes the need for explicit path sampling. As a result, the instanton method offers an efficient and tractable approach for computing rates in complex molecular systems. The rate constant is proportional to the imaginary part of the partition function, which is evaluated using Feynman's path integral formulation in Euclidean space within the steepest descent approximation. The saddle point in the path configuration space corresponds to the instanton path. To evaluate the rate, fluctuations around this path must also be included. Full technical details can be found in previous works \cite{Richardson_2018_review, kastner2014review, Beyer2016}. Here, we summarize the key results and present the instanton rate $ k_{\mathrm{inst}}$ expression:
\begin{equation}
    k_{\mathrm{inst}}(\beta) = A_{\mathrm{inst}}(\beta)\, e^{-S[\tilde{x}]/\hbar}
    \label{eq:instanton_rate}
\end{equation}
Here $\beta = 1/(k_{B}T)$ is the inverse temperature, $\tilde{x}$ is the instanton path, and $A_{\mathrm{inst}}(\beta)$ is the prefactor that accounts for the fluctuation contributions to the rate.

The implementation of the instanton algorithm involves two stages. In the first stage, the instanton path $\tilde{x}$ is located using a path optimization algorithm. In the second stage, the instanton rate $k_{\mathrm{inst}}$ is evaluated using Eq.~\ref{eq:instanton_rate}. Calculation of the prefactor $A_{\mathrm{inst}}(\beta)$ requires Hessians of the potential energy at all ring polymer beads along the instanton path, and these Hessian evaluations constitute the most computationally intensive part of the calculation.

In this work, we perform the path optimization using the efficient GPR enhanced LI-String method. The cost of the rate calculation can be reduced by training the GPR model on data from a small subset of beads and using the model to predict the Hessians of the remaining beads. Training the GPR model with Hessian information can be computationally expensive for large systems and can also be numerically unstable, an issue that has been addressed by the adaptive regression strategy previously introduced \cite{LINEB+GPR}. The selective Hessian strategy provides an additional reduction in cost by limiting explicit Hessian evaluations to a small number of beads. As an alternative, the Hessians of the ring polymer beads can be interpolated using cubic spline interpolation \cite{LI-NEB2018}. 
In numerical tests, both cubic spline interpolation method and regression can accurately approximate the rate. The cubic spline method is computationally inexpensive yet highly accurate, making it an attractive option for larger molecular systems. 

\subsection{Line Integral String Method}

In the Line Integral String (LI-String) method \cite{Cvitas2016, CvitasStringInstanton2018}, the instanton path is represented by a string, which is a smooth curve with an intrinsic parameterization. In our approach, the string is discretized into beads that are maintained at equal arc length spacing. Forces are projected perpendicular to the path to prevent the parallel component from interfering with bead redistribution. Instead of using spring forces to maintain the bead distribution, as in the LI-NEB method, we introduce a reparameterization step whenever the bead spacing becomes uneven. In our implementation, the fast inelastic relaxation engine (FIRE) method \cite{FIRE1, FIRE2} provides an efficient optimizer for locating the instanton path. Other optimizers, including quick-min and L-BFGS, can also be used \cite{NEBOptimizer}. Additional details are provided in Appendix~\ref{appendix:listring}. The LI-String method has also been applied to locate instantons at zero temperature and to compute tunneling splittings \cite{Cvitas2016, CvitasStringInstanton2018}.

The constrained dynamics approach \cite{witkin1997physically, LINEB+GPR} is used to compute the path temperature accurately. In addition, GPR can enhance the LI-String method by allowing the path optimization to proceed on a surrogate potential energy surface.

\subsection{Gaussian Process Regression}

Gaussian Process Regression (GPR) is a machine learning method capable of learning the potential energy surface (PES) of molecules and materials. It has been widely applied to accelerate geometry optimization, transition state searching \cite{Koistinen2020, Denzel2020, Jonsson2026}, and path searching \cite{Koistinen_2016, Koistinen2019}. For the present application, GPR can generate a highly accurate representation of the PES in the vicinity of the instanton path using a relatively small number of training points. Once trained, the GPR model can be used to predict gradients 
and Hessians 
at the ring polymer beads, which can accelerate the instanton rate calculation. The GPR-based optimization of the instanton pathway can be formulated within a Bayesian optimization framework, and is performed iteratively in two steps: optimizing the path on the surrogate PES generated by GPR, followed by incorporation of new \textit{ab initio} data to update the surrogate PES. Details of applying GPR to accelerate instanton path searching can be found in the related literature \cite{GPR_Instanton, Laude2020, Fang2024} and in our previous work \cite{LINEB+GPR}.

A key advantage of GPR is its ability to provide uncertainty estimates together with its predictions. By incorporating the GPR force uncertainty in the convergence criterion, the number of required force evaluations can become independent of the number of beads used to converge the instanton path \cite{LowScalingGPR2019}. The procedure for computing the force uncertainty estimate is described in Appendix~\ref{appendix:force uncertainty estimates}.

\subsubsection{GPR Hyperparameter Optimization}
Although GPR can accelerate instanton path optimization and instanton rate calculations, it also introduces non-negligible computational overhead. In particular, the optimization of GPR hyperparameters is computationally inefficient, and this bottleneck becomes more severe when Hessians are included in the GPR model. To address this challenge, we introduce an efficient GPR training protocol, which integrates  (i) physically motivated kernel lengthscale priors, (ii) Hessian-free hyperparameter optimization and (iii) Blackbox Matrix Matrix Multiplication (BBMM) accelerated by GPU hardware. This approach improves GPR model convergence and accelerates training by more than an order of magnitude.

When applying GPR method to model potentials, gradients and Hessians in high dimensional spaces, the scales of gradients and Hessians can span several orders of magnitudes across dimensions. Ignoring this anisotropy in the modeling can lead to numerical instability and difficulties in model training. To compensate for scale disparities, the kernel length-scale $l_{i}$ is chosen to be inversely proportional to the corresponding force magnitudes. Incorporating this physically motivated relations into GPR priors improves numerical stability and accelerates the model training. See Appendix \ref{Appendix:Kernel Lengtscale} for detailed discussion of the procedure.

The GPR hyperparameters include kernel lengthscales $\{ l_{i} \}$, kernel amplitude $\sigma_{m}$ and noise strength of potential $\sigma_{V}$, force $\sigma_{f,x}$ and Hessians $\sigma_{H,x}$. The performance of the GPR model is not sensitive to hyperparameters, as long as they are in a reasonable range. Training GPR model with only potential energies and gradients is sufficient to achieve convergence to appropriate hyperparameter values, with the exception of the Hessian noise parameter $\sigma_{H}^{2}$, whose optimal value can be determined straightforwardly through a small number of trail evaluations. The optimized hyperparameters are subsequently used to construct the full covariance matrix involving potential, gradient and Hessian observables. The Hessian data are still essential for accurate Hessian predictions, however, they are excluded in the model training stage to reduce the computational cost.

Furthermore, we adopt the GPU enhanced Blackbox Matrix Matrix Multiplication (BBMM) approach implemented in GPyTorch to accelerate GPR model optimization \cite{gpytorch2018}. The BBMM method reduces the cost of exact Gaussian process inference from $\mathrm{O}(n^{3})$ to $\mathrm{O}(n^{2})$. In BBMM, the optimization of GPR hyperparameters relies solely on matrix vector multiplications, which can be efficiently accelerated using parallel computing and GPUs. We briefly review the method in Appendix \ref{Appendix:BBMM} and refer the reader to Ref.~\citenum{gpytorch2018} for complete details. 
In Appendix \ref{appendix:GPU-Acceleration}, we demonstrate the improved performance of GPU-accelerated GPR training using the BBMM method.

\subsubsection{Outlier Detection Prior to GPR Modeling}
GPR is inherently sensitive to outliers because the model seeks to explain all training observations through a smooth probabilistic function. Even a small number of erroneous data points can disproportionately influence the prediction, leading to degraded model performance. Consequently, the performance of GPR, as with most machine learning methods, depends critically on the quality of the training data. To ensure robust model construction, it is therefore important to perform rigorous data validation and outlier detection prior to model training. Identifying and removing anomalous observations is essential for increasing the reliability of the resulting surrogate model. In Appendix \ref{appendix:outlier_removal}, using one representative Hessian matrix element along the instanton path of the dinitro-HBQ molecule, we demonstrate the outlier removal procedure. 

\section{Algorithmic Acceleration of Instanton Calculations}
\label{sec:algoaccel}
\subsection{Low-Scaling Line Integral String Algorithm}

By utilizing force uncertainty estimates from the GPR model, the number of force evaluations required to converge the path has been shown to be independent of the number of beads in the NEB method \cite{LowScalingGPR2019}. This allows the selection of the optimal number of beads to represent the path at no additional cost. In this section, we show that a similar advantage extends to the GPR-enhanced LI-String method.

\begin{figure}
    \centering
    \includegraphics[width=1.0\linewidth]{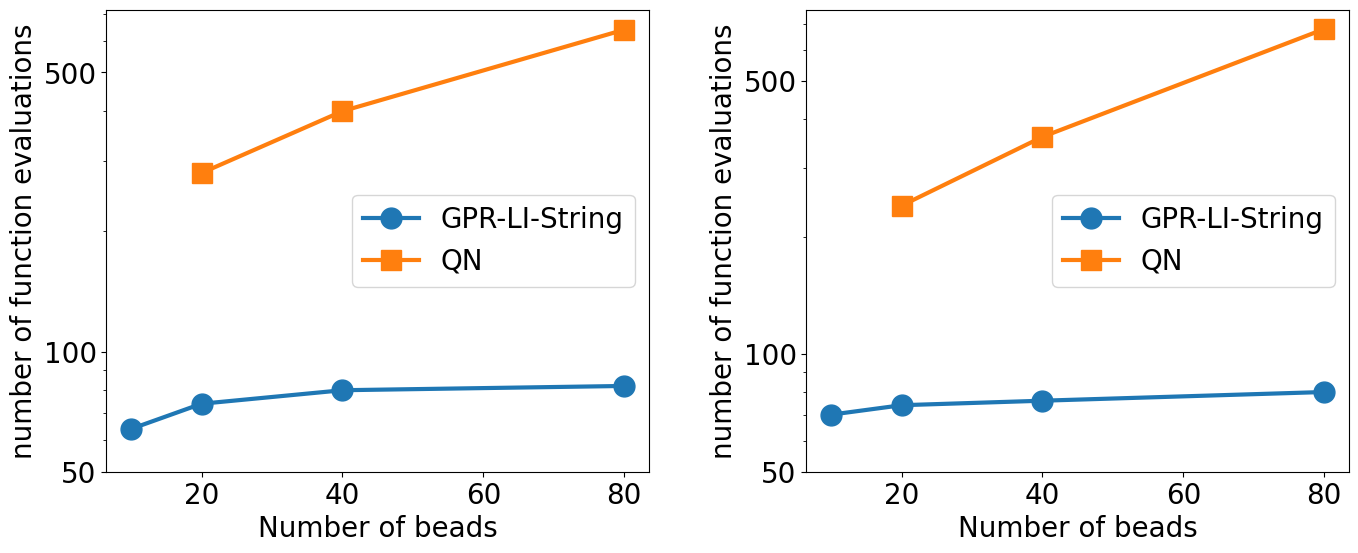}
    \caption{Comparison of the number of force evaluations required to achieve convergence as a function of the number of beads for the Quasi-Newton (QN) method and the GPR-enhanced LI-String method (GPR-LI-String). (a) Malonaldehyde instanton path at $T = 250$~K. (b) Aminopropenal instanton path at $T = 250$~K.}
    \label{fig:low_scaling_LIString}
\end{figure}

In complex molecular systems, a large number of beads are typically required to represent the path with reasonable accuracy. In the conventional instanton path searching algorithm, this is achieved by starting the path search with a small number of beads and then increasing the bead number through path interpolation followed by further path optimization. In general, increasing the number of beads leads to a larger number of required force evaluations. In contrast, by using force uncertainty estimates from the GPR model, we show in Fig.~\ref{fig:low_scaling_LIString} that the number of force evaluations becomes effectively independent of the number of beads used to represent the path in the LI-String method.

We demonstrate the advantage of the GPR-LI-String method by comparing its performance with the conventional Hessian-based instanton optimization algorithm, the Quasi-Newton (QN) method \cite{nocedal1999numerical}, using malonaldehyde and Z-3-aminopropenal as examples. In both cases, the convergence threshold for the GPR prediction force error is set to $f_{c} = 0.0025 \; \mathrm{a.u.}$. In the QN method, we first locate the instanton path using $N = 20$ beads and then further optimize the path by increasing the number of beads to achieve higher accuracy. Cubic spline interpolation is used to determine the starting geometries of the new ring polymer beads. The ring polymer Hessian expansion approach \cite{Litman2019} is employed to interpolate Hessians at the locations of the new beads, which eliminates the need for additional Hessian calculations. However, additional force evaluations are still required to optimize the path.

In Fig.~\ref{fig:low_scaling_LIString}(a), we compare the performance of the GPR-LI-String and QN methods for malonaldehyde at $T = 250$~K. Using the QN method, 280 force evaluations are required to find the instanton path with $N = 20$ beads, with an additional 120 and 240 force evaluations needed to further optimize the path with $N = 40$ and $N = 80$ beads, respectively. In contrast, using the GPR-enhanced LI-String method, 64 force evaluations are required to locate the instanton path with $N = 10$ beads, and increasing the number of beads to $N = 80$ requires only 8 additional force evaluations. This demonstrates favorable scaling of the computational cost with respect to the number of beads. 

In Fig.~\ref{fig:low_scaling_LIString}(b), we perform the benchmark test for aminopropenal at $T = 250$~K. Using the QN method, 240 force evaluations are required to locate the instanton path with $N = 20$ beads, with an additional 120 and 320 force evaluations needed to further optimize the path with $N = 40$ and $N = 80$ beads, respectively. In contrast, using the GPR-enhanced LI-String method, 70 force evaluations are required to locate the instanton path with $N = 10$ beads. Increasing the number of beads from $N = 10$ to $N = 80$ requires only 10 additional force evaluations.

\subsection{Adaptive Regression, Selective Hessian Training, and Hessian Interpolation}

In this section, we demonstrate the advantage of selective Hessian training for intramolecular proton transfer systems. The internal coordinates form a high-dimensional space, in which the instanton path spans a subspace that we refer to as the active subspace. Its orthogonal complement is the null subspace. Examining how each internal mode contributes to the tangent direction of the path reveals which modes form the active subspace and which belong to the null subspace. In proton transfer reactions, the active subspace consists of internal coordinates that are strongly coupled to the transferring proton and are referred to as flexible internal modes. The null subspace comprises coordinates that are weakly coupled to the proton, which we referred to as rigid internal modes. In Appendix \ref{appendix:flexible-rigid-modes}, we illustrate the different characteristics of flexible and rigid modes using malonaldehyde as a representative example.

The rigid modes, which lie in the null subspace, act as irrelevant input variables in the GPR model.  Furthermore, the forces and Hessians associated with rigid modes are much smaller in magnitude than those along flexible internal modes. Without proper kernel lengthscale prior, including rigid modes into the GPR model can adversely affect the hyperparameter optimization, resulting in numerical instability and poor model generalization. Fortunately, the force and Hessian associated with the rigid modes exhibit an approximately linear dependence on the internal coordinates, making them well suited for linear regression. To address this challenge, we employ an adaptive regression strategy \cite{LINEB+GPR}. The gradients and Hessians of the potential $V(x)$ in the active subspace are modeled using GPR, while those in the null subspace are modeled using linear regression. To ensure stable hyperparameter optimization and robust predictive performance, appropriate kernel length-scale priors are imposed in the GPR model. Further details are provided in Appendix \ref{Appendix:Kernel Lengtscale}. 

The partial Hessians in the null subspace are modeled using a linear regression approach, which requires fewer beads for accurate representation. In contrast, the active subspace employs a more expressive GPR model and therefore requires a larger number of beads to achieve accurate modeling. The selective Hessian training approach \cite{Fang2024}, applied separately to the active and null subspaces, leads to substantial reductions in computational cost. We illustrate this approach using ground-state proton transfer in malonaldehyde and Z-3-aminopropenal.

The rigorous instanton rate is computed using a potential energy surface generated on the fly with density functional theory (DFT). The instanton rate calculation can be accelerated by using the surrogate energy surface obtained from the adaptive regression strategy. In addition, cubic spline interpolation can be used to approximate Hessians at the ring polymer beads. For all cases presented below, we benchmark the rigorous instanton rate against the approximate rate. For the approximate rate calculations, we consider three cases that differ in the construction of the training data and the method used to approximate Hessians: (a) full Hessians are computed at selected beads for GPR modeling, (b) partial Hessians in the null subspace are computed at three beads and partial Hessians in the active subspace are computed at selected beads for GPR modeling, and (c) full Hessians are computed at selected beads for cubic spline interpolation.

\section{Applications} \label{sec:Applications}

The following examples examine systems of increasing complexity to assess the performance of the proposed framework for path optimization and rate evaluation.

\subsection{Ground state proton transfer rates of malonaldehyde}

In Table~\ref{tab:malonaldehyde_rate}, we report the instanton rate constant $k$ for ground-state intramolecular proton transfer in malonaldehyde at two temperatures, $T = 275$~K and $T = 200$~K. The ground state potential energy surface of the malonaldehyde is computed using B3LYP exchange-correlation function ~\cite{becke1988,becke1993,lee1988,vosko1980,stephens1994} and 6-31G* basis set.~\cite{hariharan1973,hehre1972} At the higher temperature, $T = 275$~K, using 10 full Hessian training data points in the model allows prediction of the instanton rate with less than $5\%$ error. If we use 3 partial Hessian data points along 13 rigid modes and 10 partial Hessian data points along 8 flexible modes as training data, the approximate instanton rate remains accurate, with an error of about $16\%$. This approach reduces the number of force evaluations by $44\%$ compared to using the full Hessian dataset. The Hessians of the ring polymer beads can also be approximated using cubic spline interpolation with 10 Hessian data points spaced equally along the instanton path, yielding an approximate instanton rate with less than $1\%$ error. 

At the lower temperature, $T = 200$~K, accurately predicting the instanton rate using the surrogate energy surface requires more training data. We find that using 3 partial Hessian data points along 13 rigid modes and 20 partial Hessian data points along 8 flexible modes as training data allows the instanton rate to be approximated within $10\%$ error relative to the rigorous instanton rate. This approach reduces the number of force evaluations by $52\%$ compared to using the full Hessian dataset. The Hessians of the ring polymer beads can also be approximated using cubic spline interpolation, resulting in an approximate rate with an error of less than $3\%$.

The numerical results demonstrate that both the cubic spline interpolation method and the Gaussian Process Regression (GPR) approach are capable of accurately reproducing the instanton rate. The cubic spline method offers advantages in terms of simplicity of implementation and computational efficiency. Nevertheless, with the improved training procedure, the computational cost of GPR model training remains modest, making it a practical and efficient alternative. The adaptive regression strategy we adopt effectively reduce the number of required Hessian calculations while maintaining the accuracy in the instanton rate calculation.
\begin{table}[]
    \centering
    \resizebox{0.8 \textwidth}{!}{
    \begin{tabular}{|c|c|c|c|c|}
    \hline
     T(K)  &  Training set & bead number N & rate constant ($ps^{-1}$) & error ($\%$) \\ \hline 
         \multirow{8}{*}{275} 
                    & (a) 20V, 20 G, 10 H & 40 & $6.30$ & $3.2$ \\ 
                    &                     & 320 & 6.57 &   $3.4$ \\
        \cline{2-5}
                    & (b) 20V, 20 G & 40 & 5.44 & 16.4 \\ 
                    &   10 H (8 flexible), 3 H (13 rigid)              & 320 & 5.68 &   16.4 \\
        \cline{2-5}
                    & (c) 20 H (cubic spline) & 40 & 6.52 &  0.1 \\ 
                    &                 & 320 & 6.79 &   0.1 \\
        \cline{2-5}
                    & (d) DFT              &  40 & 6.51 &   \\
                    &                  & 320 & 6.80 &  \\
        
        \hline 
         \multirow{6}{*}{200} 
                    & (a) 30V, 30 G, 20 H & 40  & 3.19 & 10.3 \\ 
                    &                 & 320 & 3.47 &  11.7  \\
        \cline{2-5}
                    & (b) 30V, 30 G, 20 H 
                                            & 40 & 3.23 & 9.3 \\ 
                    &  (8 flexible), 3 H (13 rigid)
                    & 320 & 3.56 & 9.4   \\
        \cline{2-5}
                    & (c) 20 H (cubic spline)              &  40 & 3.47 &  2.5 \\ 
                    &                  & 320 & 3.83 & 2.5 \\
        \cline{2-5}
                    & (d) DFT              &  40 & 3.56 &   \\
                    &                  & 320 & 3.93 &  \\
        \hline         
    \end{tabular}
    }
    \caption{Ground state intramolecular proton transfer in malonaldehyde. Comparison of the instanton rate computed (a) on a surrogate potential energy surface, where Hessians in the active subspace are fitted using GPR, employing the same number of beads as used for the null subspace, which is modeled with linear regression. (b) On a surrogate potential energy surface, where Hessians in the null subspace are fitted with only 3 beads. (c) Hessians are fitted with cubic spline interpolation. (d) Potential energy surface computed using DFT. For malonaldehyde, there are 21 internal modes.}
    \label{tab:malonaldehyde_rate}
\end{table}

\subsection{Ground state proton transfer rate of Z-3-aminopropenal}
In Table~\ref{tab:aminopropenal-rate}, we report the instanton rate constant $k$ for ground-state intramolecular proton transfer in Z-3-aminopropenal at two temperatures, $T = 250$~K and $T = 160$~K. The ground state potential energy surface of the Z3-aminopropenal is computed using B3LYP exchange-correlation function ~\cite{becke1988,becke1993,lee1988,vosko1980,stephens1994} and 6-31G* basis set.~\cite{hariharan1973,hehre1972} At $T = 250$~K, using 3 partial Hessian training data points along rigid modes and 20 partial Hessian training data points along flexible modes, the instanton rate predicted on the surrogate surface is within $15\%$ error relative to the rigorous rate. Along the instanton path at $T = 250$~K, there are 12 flexible modes and 12 rigid modes. This approach reduces the number of force evaluations by $35\%$ compared to using the full Hessian dataset. We find that the approximate instanton rate calculated with Hessians interpolated by cubic spline interpolation is also highly accurate, with an error within 3 $\%$.
At the lower temperature, $T = 160$~K, the instanton rate predicted using partial Hessian data has an error of approximately $20\%$ relative to the rigorous rate. Along the instanton path at $T = 160$~K, there are 12 flexible modes and 12 rigid modes. This approach reduces the number of force evaluations by $42\%$ compared to using the full Hessian dataset. The approximate rate can also be calculated with Hessians interpolated by cubic spline interpolation, and we find that the error in the rate is around $3\%$. 

Both cubic spline interpolation and GPR accurately approximate instanton rates, provided that appropriate data preprocessing is performed, including outlier removal. While GPR requires careful treatment of flexible internal modes, spline interpolation offers a simpler and more computationally efficient alternative for instanton rate calculations.


\begin{table}[]
    \centering
    \resizebox{0.8 \textwidth}{!}{
    \begin{tabular}{|c|c|c|c|c|}
    \hline
     T(K)   & Training set & bead number N & rate ($ps^{-1}$) & error ($\%$) \\ \hline 
         \multirow{8}{*}{250} & (a) 80V, 80 G, 20 H & 40 & 155.5 & 4.8 \\ 
                         &               & 320 & 179.1 &   15.2 \\
        \cline{2-5}
                         &(b) 80 V, 80 G
                         & 40 & 169.8 & 14.4 \\
                         &     20 H (12 flexible), 3 H (12 rigid)             & 320 & 173.4 & 11.5 \\
        \cline {2-5}     
                         & (c) 20 H (cubic spline)              &  40 & 145.3 &  2.1 \\
                         &               & 320 & 151.6 & 2.4 \\
        \cline {2-5}     
                         &(d)DFT              &  40 & 148.4 &   \\
                         &                & 320 & 155.4 &  \\
        
        \hline
        \multirow{8}{*}{160} &  (a) 80 V, 80 G, 20 H & 40 & 99.8 & 9.7 \\
                         &                 & 320 & $112.0$ & $6.1$ \\
        \cline{2-5}
                         & (b) 80 V, 80 G       & 40 & $110.7$ & $21.6$ \\
                         &   20 H (12 flexible) , 3 H (12 rigid)              & 320 & $123.5$ & $17.1$ \\
        \cline {2-5}     
                        &(c) 20 H (cubic spline)               &  40 & 93.8 & 3.1  \\
                        &               & 320 & 105.5 & 0.1 \\
        \cline {2-5}     
                        &(d) DFT               &  40 & 91.0 &   \\
                         &                & 320 & 105.5 &  \\
        \cline{2-5}
        \hline
    \end{tabular}
    }
    \caption{Ground state intramolecular proton transfer in Z3-aminopropenal. Comparison of the instanton rate computed (a) on a surrogate potential energy surface, where Hessians in the active subspace are fitted using GPR, employing the same number of beads as used for the null subspace, which is modeled with linear regression. (b) On a surrogate potential energy surface, where Hessians in the null subspace are fitted with only 3 beads. (c) With Hessians fitted with cubic spline interpolation. (d) Potential energy surface computed using DFT. For aminopropenal, there are 24 internal modes.}
    \label{tab:aminopropenal-rate}
\end{table}

\subsection{Ground state proton transfer rate of 7,9-dinitro-HBQ}
In this section, we report the ground state tunneling rate constant $k$ for the dinitro-HBQ molecule, computed using the instanton method. The ground state potential energy surface is computed using DFT with the PBE0~\cite{adamo1999} 
exchange-correlation functional and 6-31G* basis set ~\cite{hariharan1973,hehre1972}. The ground state barrier is $\sim$1.15 kcal/mol, which lies in the range accessible by thermal excitation at room temperature. This means the tunneling dynamics here is shallow tunneling. The dinitro-HBQ is larger than the two molecules previously examined in this work. 
Compared with malonaldehyde and Z-3-aminopropenal,
the two nitro group substituents introduce low-frequency torsion modes, further complicating the machine learning approximation of the instanton rate, because the predicted Hessians must be highly accurate to prevent the contamination of these low frequency modes.  We note that the zero-frequency mode associated with the instanton pathway may acquire a finite frequency as a consequence of numerical errors introduced by the machine-learned potential. 
The existence of low-frequency torsional modes further complicates the identification of the true zero mode based solely on the computed eigenvalue spectrum. However, the eigenvector corresponding to the zero mode is expected to exhibit substantial overlap with the motion of the transferring proton. Consequently, the true zero mode can be reliably identified through examination of the associated eigenvectors rather than by considering the eigenvalues alone.

Because the ground state barrier is low, the tunneling rate constants at two different temperatures considered are close to each other. Another factor that can substantially influence the computed instanton rate is the limited accuracy of the conventional instanton theory for shallow tunneling close to barrier top. This deficiency can be mitigated by the recently proposed microcanonical instanton theory. \cite{microcanonicalinstanton1, microcanonicalinstanton2}. 
However, as this is beyond the scope of the present work, we do not employ this approach and instead note it as a potential source of error in the reference instanton rates.

The dinitro-HBQ system also highlights the increasing challenges of applying global GPR models to larger molecular systems. The GPR approach becomes more computationally demanding because the cost and memory requirements increase rapidly with dimensionality, particularly when Hessian information is included. To mitigate this, we apply an adaptive regression strategy that retains only 40 of the 78 internal modes in the GPR model and further improve the efficiency of model training using Hessian-free optimization, a physics-informed kernel prior, and the GPU-accelerated BBMM approach. These measures make Hessian-enabled GPR calculations computationally tractable for the present system, while also illustrating the challenges associated with constructing accurate global surrogate models for higher-dimensional problems.


The instanton rate constant of dinitro-HBQ is shown in Table \ref{tab:dinitro-HBQ-rate}. 
The GPR-predicted instanton rates still show a noticable deviation from the \textit{ab initio} results. We attribute this discrepancy primarily to the omission of some nonlinear modes from the GPR model. Although incorporating additional degree of freedoms in the GPR model will likely improve the rate predictions, this is currently precluded by memory constraints: in our present implementation, the kernel matrix storage already approaches the limit of our GPU hardware. 

Our numerical experiments on dinitro-HBQ show that cubic-spline interpolation yields a computationally inexpensive yet highly accurate approximation to the instanton rate, particularly for larger molecular systems. The favorable performance can be attributed to the local nature of spline interpolation, which relies only on neighboring data points. By comparison, GPR employs a global regression framework that incorporates information from all training points along the instanton pathway. While this global treatment offers considerable flexibility, it also results in computational costs that increase with the size of training dataset and molecular system.  

\begin{table}[]
    \centering
    \resizebox{0.8 \textwidth}{!}{
    \begin{tabular}{|c|c|c|c|c|}
    \hline
     T(K)    & Training set & bead number N & rate ($ps^{-1}$) & error ($\%$) \\ \hline 
         \multirow{6}{*}{193} & (a) 30V, 30 G, 20 H & 40 & 474.6 & 40.5 \\ 
                         &               & 320 & 504.9 &  37.4  \\
        \cline {2-5}     
                         & (b) 20 H (cubic spline)  &  40 & 337.8 & 0.1 \\
                         &               & 320 & 365.2 & 0.6 \\
        \cline {2-5}     
                         &(c)DFT              &  40 & 337.6 &   \\
                         &                & 320 & 367.4 &  \\
        
        \hline
        \multirow{6}{*}{137} &  (a) 30 V, 30 G, 20 H & 40 & 104.7 & 69.4  \\
                         &                 & 320 & 122.5 & 70.1\\
        \cline {2-5}     
                        &(b) 20 H (cubic spline)     &  40 &  354.4 & 3.4 \\
                        &               & 320 & 419.2  & 2.1 \\
        \cline {2-5}     
                        &(c) DFT               &  40 & 342.7 &   \\
                        &                & 320 &  410.6 &  \\
        \cline{2-5}
        \hline
    \end{tabular}
    }
    \caption{Ground state intramolecular proton transfer in dinitro-HBQ. Comparison of the instanton rate computed (a) on a surrogate potential energy surface, where Hessians in the active subspace are fitted using GPR, employing the same number of beads as used for the null subspace, which is modeled with linear regression. (b) with Hessians fitted with cubic spline interpolation. (c) potential energy surface computed using DFT. For dinitro-HBQ, there are 28 atoms and 78 internal modes.}
    \label{tab:dinitro-HBQ-rate}
\end{table}


\section{Conclusions}
\label{sec:conclusions}

In this work, we developed a GPR-enhanced Line Integral String (LI-String) method to accelerate ring polymer instanton calculations. By leveraging the GPR uncertainty quantification, we show that the cost of converging the instanton path no longer scales with the number of beads representing the path. A drawback of the traditional GPR implementation is that the computational overhead associated with model training is non-negligible, which limits applicability to larger systems, especially when Hessians are included in the training data. Conventional GPR training is inefficient and exhibits unfavorable cubic scaling with the number of training data points, $\sim \mathcal{O}(N^{3})$. Here, we implement an efficient GPR training protocol, integrating the proper GPR prior, Hessian-free model training and the Blackbox Matrix Matrix Multiplication (BBMM) method with GPU acceleration \cite{gpytorch2018} to speed up model training. This provides an order-of-magnitude speedup in GPR model training.

Once the instanton path is optimized, the remaining cost of the instanton rate calculation is still approximately $n$ times that of transition state theory, where $n$ is the number of ring polymer beads for which Hessians are required. The GPR method can construct a surrogate energy surface using a relatively small number of Hessian data points and still yield accurate instanton rates, thereby reducing the cost of Hessian evaluations. For proton transfer reactions, we further suggest that Hessian costs can be reduced by combining selective Hessian training \cite{Fang2024} with the adaptive regression strategy. This is enabled by evaluating Hessian components along rigid modes using fewer beads than are needed for the flexible modes. We illustrate this idea for malonaldehyde and Z-3-aminopropenal, where Hessians along the rigid modes are evaluated using 3 beads and Hessians along the flexible modes are evaluated using 20 beads. The predicted instanton rates remain within approximately 20\% of the ab-initio instanton rates, with reductions in the number of force evaluations ranging from 35\% to 52\%.
In addition, we show that cubic spline interpolation can also accurately approximate instanton rates using only a limited number of Hessian evaluations along the instanton path. We emphasize the importance of the detecting and removing anomalous Hessian data prior to applying the GPR method or cubic spline interpolation method, as the model performance depends critically on the data quality. 

We note that while both GPR and spline interpolation methods can yield accurate instanton rate results, their strengths are complementary: GPR is well suited for surrogate modeling and Bayesian optimization owing to its built-in uncertainty quantification and robust predictive capabilities, whereas spline interpolation offers a computationally efficient and accurate approach for accelerating instanton rate calculations once the instanton pathway has been determined. For the present application, we find that an effective strategy is to employ GPR to accelerate the instanton path optimization, while using spline interpolation to reduce the cost of Hessian evaluations in instanton rate calculation. 

For large molecular systems, evaluating the fluctuation factor requires diagonalization of the ring polymer Hessian matrix, the computational cost of which can become prohibitive. To address this challenge, a variety of efficient algorithms have been developed to reduce the cost associated with the matrix diagonalization. \cite{divide_conquer2019, McConnell2017, microcanonicalinstanton2} The instanton rate calculation depends only on the fluctuation factor that determined by the Hessian eigenvalue spectrum and is independent of the corresponding eigenvectors. This observation suggests the explicit eigendecomposition may be avoided by employing matrix-free techniques that access spectral information through matrix-vector products alone. In particular, Krylov subspace methods, such as the Lanczos algorithm, together with stochastic trace estimation techniques\cite{Hutchinson01011990, Saad2017}, provides a promising framework for efficiently evaluating the required spectral quantities. The development and application of such approaches will be the subject of the future work.

\begin{appendices}

\section{Line Integral String Optimization} 
\label{appendix:listring}

The objective function for the LI-String method is the abbreviated action $W(\mathbf{r})$:
\begin{equation}
    W(\mathbf{r}) = \frac{1}{\hbar} \int_{\mathbf{r}_{1}}^{\mathbf{r}_{n}} \sqrt{2 \left(V(\mathbf{r}) - E\right)} \, dr,
    \label{abbreviated-action-W-continuous}
\end{equation}
where $\mathbf{r}_{1}$ and $\mathbf{r}_{n}$ are the turning points.

Using mass-scaled Cartesian coordinates, the action $W(\mathbf{r})$ is discretized using $n$ beads:
\begin{equation}
W(\mathbf{r}) \approx \frac{1}{2 \hbar} \sum_{j=1}^{n} \left( \sqrt{2 \left(V(\mathbf{r}_{j}) - E\right)} + \sqrt{2 \left(V(\mathbf{r}_{j-1}) - E\right)} \right) \left|\mathbf{r}_{j} - \mathbf{r}_{j-1}\right|.
\end{equation}
The LI-String optimization is performed with constraints. The interior beads are maintained at equal arc length spacing, and the two end beads are constrained to move on the energy contour $V(\mathbf{r}) = E$.

Using Lagrange multipliers, we formulate the constrained objective function as
\begin{equation}
    L(\mathbf{r}_{1},\mathbf{r}_{2}, \cdots, \mathbf{r}_{n})
    = W(\mathbf{r}_{1},\mathbf{r}_{2}, \cdots, \mathbf{r}_{n})
    - \lambda_{1} \left(V(\mathbf{r}_{1}) - E\right)
    - \lambda_{2} \left(V(\mathbf{r}_{n}) - E\right)
    \label{eq:lagrangian_multiplier}
\end{equation}
where $\lambda_{1}$, $\lambda_{2}$ are Lagrange multipliers associated with the constraints on the two end beads respectively.

The optimization force $\mathbf{g}_{j} = - \nabla_{j} W$ is given in Ref.~\citenum{LI-NEB2018} as:
\begin{equation}
\mathbf{g}_j= -\nabla_j W = \frac{1}{2}\left(\frac{1}{\xi_j \hbar^2}(d_j + d_{j+1}) \mathbf{f}_j - (\xi_j + \xi_{j-1}) \widehat{\mathbf{d}}_j + (\xi_{j+1} + \xi_j) \widehat{\mathbf{d}}_{j+1} \right),
\label{negative-gradient-W}
\end{equation}
where $\textbf{f}_j$ is the physical force, $\xi_i = \frac{1}{\hbar} \sqrt{2 (V(\mathbf{r}_i) - E)}$, and
\begin{equation}
    \widehat{\mathbf{d}}_j = \left(\mathbf{r}_j - \mathbf{r}_{j-1} \right) / d_j, \quad \text{with} \quad d_j = \left| \mathbf{r}_j - \mathbf{r}_{j-1} \right|
\end{equation}
have been used. 

The LI-String method uses projection and reparameterization to efficiently impose the equal arc length constraint $r \, \hat{\mathbf{t}}$ for the interior beads. The optimization force for the interior bead $j$ includes only the transverse component of $\mathbf{g}_{j}$:
\begin{equation}
    \mathbf{g}_{j}^{\mathrm{opt}} = \mathbf{g}_{j}^{\perp}.
    \label{eq:transverse_opt_original}
\end{equation}
A reparameterization is performed when the bead distribution becomes uneven to re-enforce the constraint. The criterion used for repositioning the beads is
\begin{equation}
    \max_{i} \left\| \left|\mathbf{x}_{i+1}-\mathbf{x}_i\right| - \left|\mathbf{x}_i-\mathbf{x}_{i-1}\right| \right\|
    > \frac{\sum_{i}\left|\mathbf{x}_{i+1}-\mathbf{x}_i\right|}{10(n-1)}.
\end{equation}
We use cubic spline interpolation of the path to reposition the beads so that they are equally spaced along the curve.

From the Lagrangian formulation in Eq.~\ref{eq:lagrangian_multiplier}, the converged path satisfies constraints on the potential gradients and energies of the two end beads:
\begin{equation}
    \begin{aligned}
     & \sqrt{2 \mu \left(V(\mathbf{r}_{2}) - E\right)}
     \frac{\mathbf{r}_{1} - \mathbf{r}_{2}}{\left|\mathbf{r}_{1} - \mathbf{r}_{2}\right|}
     = \lambda_{1} \nabla V(\mathbf{r}_{1}), \\
     & \sqrt{2 \mu \left(V(\mathbf{r}_{n-1}) - E\right)}
     \frac{\mathbf{r}_{n-1} - \mathbf{r}_{n}}{\left|\mathbf{r}_{n-1} - \mathbf{r}_{n}\right|}
     = \lambda_{2} \nabla V(\mathbf{r}_{n}),
    \end{aligned}
    \label{eq: potential-gradient-constraint}
\end{equation}
and
\begin{equation}
    \begin{aligned}
    & V(\mathbf{r}_{1}) = E, \\
    & V(\mathbf{r}_{n}) = E.
    \end{aligned}
    \label{eq: energy-constraint}
\end{equation}

Two energy-constraint terms are introduced to constrain the end beads to the energy contour. In addition, two auxiliary force terms $\mathbf{g}_{1}^{\mathrm{aux}}$ and $\mathbf{g}_{n}^{\mathrm{aux}}$ are introduced, with magnitudes comparable to $\mathbf{g}_{j}^{\perp}$, to facilitate rapid convergence \cite{Free-end-NEB2016}:
\begin{equation}
\begin{aligned}
      & \mathbf{g}_{1}^{\mathrm{aux}} = \left\langle \left|\mathbf{g}_{j}^{\perp}\right| \right\rangle
      \frac{\mathbf{r}_{2} - \mathbf{r}_{1}}{\left|\mathbf{r}_{1} - \mathbf{r}_{2}\right|}, \\
      & \mathbf{g}_{n}^{\mathrm{aux}} = \left\langle \left|\mathbf{g}_{j}^{\perp}\right| \right\rangle
      \frac{\mathbf{r}_{n-1} - \mathbf{r}_{n}}{\left|\mathbf{r}_{n-1} - \mathbf{r}_{n}\right|}.
\end{aligned}
\end{equation}

The final optimization forces for the two end beads are
\begin{equation}
\mathbf{g}_{1,n}^{\mathrm{opt}}
=
\mathbf{g}_{1,n}^{\mathrm{aux}}
-\left(\mathbf{g}_{1,n}^{\mathrm{aux}} \cdot \hat{\mathbf{f}}\left(\mathbf{r}_{1,n}\right)\right)
\hat{\mathbf{f}}\left(\mathbf{r}_{1,n}\right)
+\kappa\left(V\left(\mathbf{r}_{1,n}\right)-E\right)\hat{\mathbf{f}}\left(\mathbf{r}_{1,n}\right),
\end{equation}
where the first two terms enforce the potential-gradient constraint in Eq.~\ref{eq: potential-gradient-constraint}, and the third term enforces the energy constraint in Eq.~\ref{eq: energy-constraint}.

The LI-String algorithm is considered converged when
\begin{equation}
|\mathbf{g}_{j}^{\mathrm{opt}}| < \epsilon^{\mathrm{interior}}, \quad
|\mathbf{g}_{1,n}^{\mathrm{opt}}| < \epsilon^{\mathrm{end}}, \quad
\sum_{j=2}^{n-1} |\mathbf{g}_{j}^{\perp}| < \epsilon^{\mathrm{sum}}.
\label{eq:LIString_convergence_criterion}
\end{equation}

At convergence, we use constrained dynamics \cite{witkin1997physically, LINEB+GPR} to define the temperature as $T = \frac{\hbar}{k_{B}\tau}$, where $k_{B}$ is the Boltzmann constant and $\tau$ is the oscillation period. This approach provides an accurate temperature estimate, particularly when the instanton path exhibits significant curvature at low temperatures.

We provide additional remarks on the transverse component of the action force $g_{j}^{\perp}$ and its impact on the convergence rate, as well as the precision of the resulting path. Maintaining a smooth variation of the transverse component of the action force $g_{j}^{\perp}$ across all images is important for ensuring consistent lateral motion of the string. \cite{Free-end-NEB2016} As seen from the expression for the action force $g_{j}$ (eq.\ref{negative-gradient-W}), the first term indicates that its magnitude near the endpoints of the path can be substantially larger than that on internal beads. To improve the precision of the resulting path, we apply a local averaging procedure for $|g_{j}^{\perp}|$ over three neighboring beads, thereby smoothing its variation along the path. This is equivalent to changing eq.\ref{eq:transverse_opt_original} to the following equation:
\begin{equation}
    g_{j}^{opt} = \hat{g_{j}^{\perp}} \frac{|g_{j-1}^{\perp}| + |g_{j}^{\perp}| + |g_{j+1}^{\perp}|}{3}
\end{equation}
Since the modified optimization gradient shares the same convergence criterion as the original, both converge to the same instanton path. Its smoother variation, however, results in faster convergence and improved accuracy.

\section{Force Uncertainty Estimation in Gaussian Process Regression}
\label{appendix:force uncertainty estimates}
To incorporate derivative information in GPR, the observable vector $\mathbf{y}$ is extended as
\begin{equation}
    \mathbf{y}_{\mathrm{ext}}
    =
    \left[
    y^{(1)} \cdots y^{(N)},
    \frac{\partial y^{(1)}}{\partial q_1^{(1)}} \cdots \frac{\partial y^{(N)}}{\partial q_1^{(N)}},
    \frac{\partial y^{(1)}}{\partial q_2^{(1)}} \cdots \frac{\partial y^{(N)}}{\partial q_2^{(N)}},
    \ldots,
    \frac{\partial y^{(1)}}{\partial q_D^{(1)}} \cdots \frac{\partial y^{(N)}}{\partial q_D^{(N)}}
    \right]^{\top}.
\end{equation}
where $N$ is the number of training data points and $D$ is the dimension. The covariance matrix $K(Q,Q)$ is extended to include covariance terms between partial derivatives and between the function values and partial derivatives:
\begin{equation}
\mathbf{K}_{\mathrm{ext}}=
\left[
\begin{array}{ccccc}
K(\mathbf{Q}, \mathbf{Q}) & \frac{\partial K\left(\mathbf{Q}, \mathbf{Q}^{\prime}\right)}{\partial q_1^{\prime}} & \frac{\partial K\left(\mathbf{Q}, \mathbf{Q}^{\prime}\right)}{\partial q_2^{\prime}} & \cdots & \frac{\partial K\left(\mathbf{Q}, \mathbf{Q}^{\prime}\right)}{\partial q_D^{\prime}} \\
\frac{\partial K\left(\mathbf{Q}, \mathbf{Q}^{\prime}\right)}{\partial q_1} & \frac{\partial^2 K\left(\mathbf{Q}, \mathbf{Q}^{\prime}\right)}{\partial q_1 \partial q_1^{\prime}} & \frac{\partial^2 K\left(\mathbf{Q}, \mathbf{Q}^{\prime}\right)}{\partial q_1 \partial q_2^{\prime}} & \cdots & \frac{\partial^2 K\left(\mathbf{Q}, \mathbf{Q}^{\prime}\right)}{\partial q_1 \partial q_D^{\prime}} \\
\frac{\partial K\left(\mathbf{Q}, \mathbf{Q}^{\prime}\right)}{\partial q_2} & \frac{\partial^2 K\left(\mathbf{Q}, \mathbf{Q}^{\prime}\right)}{\partial q_2 \partial q_1^{\prime}} & \frac{\partial^2 K\left(\mathbf{Q}, \mathbf{Q}^{\prime}\right)}{\partial q_2 \partial q_2^{\prime}} & \cdots & \frac{\partial^2 K\left(\mathbf{Q}, \mathbf{Q}^{\prime}\right)}{\partial q_2 \partial q_D^{\prime}} \\
\vdots & \vdots & \vdots & \ddots & \vdots \\
\frac{\partial K\left(\mathbf{Q}, \mathbf{Q}^{\prime}\right)}{\partial q_D} & \frac{\partial^2 K\left(\mathbf{Q}, \mathbf{Q}^{\prime}\right)}{\partial q_D \partial q_1^{\prime}} & \frac{\partial^2 K\left(\mathbf{Q}, \mathbf{Q}^{\prime}\right)}{\partial q_D \partial q_2^{\prime}} & \cdots & \frac{\partial^2 K\left(\mathbf{Q}, \mathbf{Q}^{\prime}\right)}{\partial q_D \partial q_D^{\prime}}
\end{array}
\right].
\end{equation}
The variance of the posterior distribution for the potential and force at an unknown point in internal coordinates $q^{*}$ (corresponding to Cartesian coordinates $x^{*}$) is
\begin{equation}
    \Lambda^{\mathrm{pred}}_{qq}
    =
    K_{\mathrm{ext}}(q^{*}, q^{*})
    - K_{\mathrm{ext}}(q^{*}, Q)\left(K_{\mathrm{ext}}(Q,Q) + \Lambda_{QQ}\right)^{-1}K_{\mathrm{ext}}(Q, q^{*}).
\end{equation}
The variance of the posterior distribution in Cartesian coordinates can be obtained using the transformation matrix $L$ \cite{Fang2024}:
\begin{equation}
    \begin{aligned}
        \Lambda_{xx}^{\mathrm{pred}}(x^{*})
        &=
        \left(
        \begin{array}{cc}
        \mathrm{Var}(V) & \mathrm{Cov}(V,f_{x}) \\
        \mathrm{Cov}(f_{x},V) & \mathrm{Var}(f_{x})
        \end{array}
        \right)
        =
        L\, \Lambda^{\mathrm{pred}}_{qq}(q^{*})\, L^{T}, \\
        L
        &=
        \left(
        \begin{array}{cc}
        1 & 0 \\
        0 & B_{q}^{T}
        \end{array}
        \right),
    \end{aligned}
\end{equation}
where $B_{q} = \frac{dq}{dx}$ is Wilson's $B$ matrix. The force uncertainty at the unknown point $x^{*}$ can be estimated by the trace of the force variance matrix $\mathrm{Var}(f_{x})$:
\begin{equation}
    \sigma_{f,x}^{\mathrm{pred}}(x^{*})
    =
    \sqrt{\mathrm{Tr}\left(\mathrm{Var}\left(f_{x}(x^{*}) \right) \right)}.
\end{equation}

\section{Constraints on GPR kernel length scales}
\label{Appendix:Kernel Lengtscale}
In our previous work \cite{LINEB+GPR}, we highlighted the importance of properly regularizing hyperparameters to ensure successful GPR model training. Here we emphasize the importance to constrain kernel length scales, particularly when modeling Hessian data in high dimensional potential energy surface. Force and Hessian magnitudes vary across dimensions, indicating that the kernel lengthscales should also differ by dimensions. Providing this lengthscale information to the GPR model can improve model training and improve overall performance. We put constraints on kernel length scales by defining them as the transformation of raw length scales $l_{k}^{\mathrm{raw}}$, which are treated as trainable hyperparameters.
\begin{equation}
    \begin{aligned}
        l_{k} = \sigma(l_{k}^{\textrm{raw}}) (U-L) + L \\
        \sigma(x) = \frac{1}{1+ e^{-x}}
    \end{aligned}
\end{equation}
In our implementation, we compute average force amplitude along different dimensions and set lengthscale limit as the inverse of average force amplitude ($L = 0.5/\langle f \rangle$, $U = 5 / \langle f \rangle$).

\section{Efficient GPR model training strategy} \label{Appendix:BBMM}
The performance of the GPR model depends on a set of hyperparameters $\theta$, which include likelihood noise and kernel lengthscales $l_{d}$. During model training, these hyperparameters are optimized by minimizing the negative log marginal likelihood,
\begin{equation}
    L(\theta|X,y)
    = -\log p(\theta|X,y)
    = \frac{1}{2} y^{T} K_{XX}^{-1} y 
      + \frac{1}{2} \log(|K_{XX}|)
      + \frac{n}{2} \log(2\pi).
\end{equation}

In the conventional approach, Cholesky decomposition is used to compute $K_{XX}^{-1}y$ and the log determinant $\log(|K_{XX}|)$. The $\mathrm{O}(n^{3})$ cost of the Cholesky decomposition makes this approach inefficient for large training sets. In the BBMM method \cite{gpytorch2018}, the linear conjugate gradient algorithm \cite{nocedal1999numerical} is used to solve $K_{XX}^{-1}y$, and the trace estimator \cite{Hutchinson01011990, Hutch++2021} is used to approximate $\log(|K_{XX}|)$. The algorithm integrates the Lanczos tridiagonalization procedure required for log determinant evaluation with the conjugate gradient iterations and employs a preconditioner to accelerate convergence.

In practice, the conjugate gradient method converges in $p$ iterations and has computational cost $\mathrm{O}(pn^{2})$ with $p \ll n$. The Hutchinson trace estimator \cite{Hutchinson01011990} approximates $\log(|K_{XX}|)$ using $t$ random vectors with computational cost $\mathrm{O}(tn^{2})$, where $t \ll n$. Together, these ingredients reduce the overall time complexity of GP inference to $\mathrm{O}(n^{2})$, which is more efficient than the $\mathrm{O}(n^{3})$ cost associated with the Cholesky decomposition.

The performance of the GPR model is not highly sensitive to the choice of hyperparameters. Hyperparameters that yield a good fit to the potential energy and forces typically also provide an accurate fit to the Hessian. In our numerical tests, we optimize the hyperparameters using only potential and gradient data. This approach reduces the computational cost of hyperparameter optimization, since a smaller covariance matrix needs to be constructed and inverted. After training, the optimized hyperparameters are then used to build the covariance matrix over the full dataset, including potential, force and Hessian data. We find this optimization strategy effective, as the GPR-predicted Hessian errors remain small.

\section{GPU-Accelerated GPR Hyperparameter Optimization}
\label{appendix:GPU-Acceleration}
To demonstrate the improved performance of the GPU-accelerated BBMM method for training the GPR model, we benchmark it against BBMM on CPU and GPU, the Cholesky method on CPU and GPU using
three example molecules: malonaldehyde, Z-3-aminopropenal, and dinitro-HBQ. 
For both the Cholesky algorithm and BBMM method, we use the implementation provided in GPyTorch. The training data for malonaldehyde include 20 potential and gradient data points and 10 Hessian data points. Among the 21 internal modes, 8 flexible modes are included in the GPR model, and the remaining 13 rigid modes are modeled by linear regression.
The training data for Z-3-aminopropenal include 80 potential and gradient data points and 20 Hessian data points. Among the 24 internal modes, 12 modes are included in the GPR model, and the remaining 12 modes are modeled by linear regression. 
The training data for dinitro-HBQ include 23 potential and gradient data points and 20 Hessian data points. Among the 78 internal modes, 15 modes are included in the GPR model, and the remaining 63 modes are modeled by the linear regression.

\begin{figure}[H]
    \centering
    \includegraphics[width=0.6\linewidth]
    {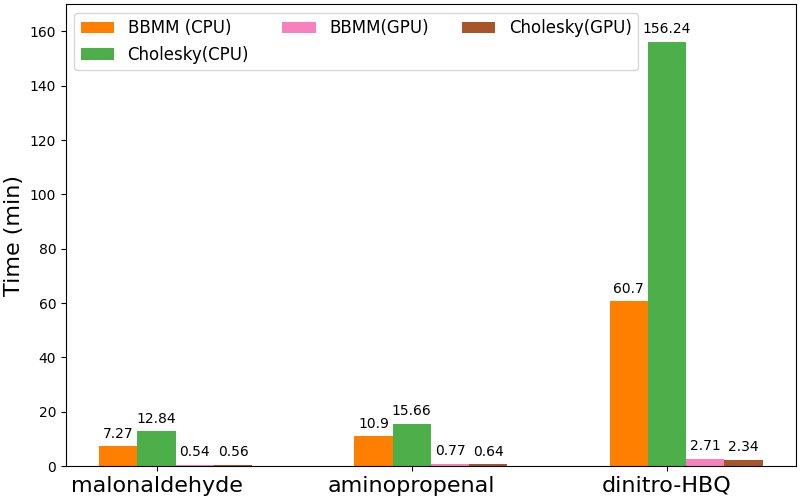}

    \caption{GPR hyperparameter training time for three different molecules using the  BBMM on CPU (orange), Cholesky decomposition on CPU (green), BBMM on GPU (pink), and Cholesky decomposition on GPU (brown), reported in minutes. Details of the training datasets are provided in the text. 
    }
    \label{fig:GPR_training_time}
\end{figure}

The GPR model training times are shown in Fig.~\ref{fig:GPR_training_time}. Both the GPU-accelerated BBMM method and the GPU-accelerated Cholesky method are efficient for GPR training. For all three example molecules, GPR model training on GPU is completed within a few minutes. 
Although the theoretical time complexity of the BBMM method is lower than that of the Cholesky algorithm, this advantage is not significant for the test cases studied here. Compared to the GPR model trained with the BBMM method on CPU, the GPU implementation is an order of magnitude faster for the molecular systems we have considered, demonstrating the suitability of the BBMM method for GPU acceleration.
The Cholesky method on CPU is the most computationally inefficient implementation among all methods tested. Notably, in our tests, the Cholesky method exhibits a larger performance improvement upon GPU acceleration than the BBMM method.

\section{Outlier Removal Procedure}
\label{appendix:outlier_removal}
\begin{figure}[H]
    \centering
    \includegraphics[width=0.8\linewidth]
    {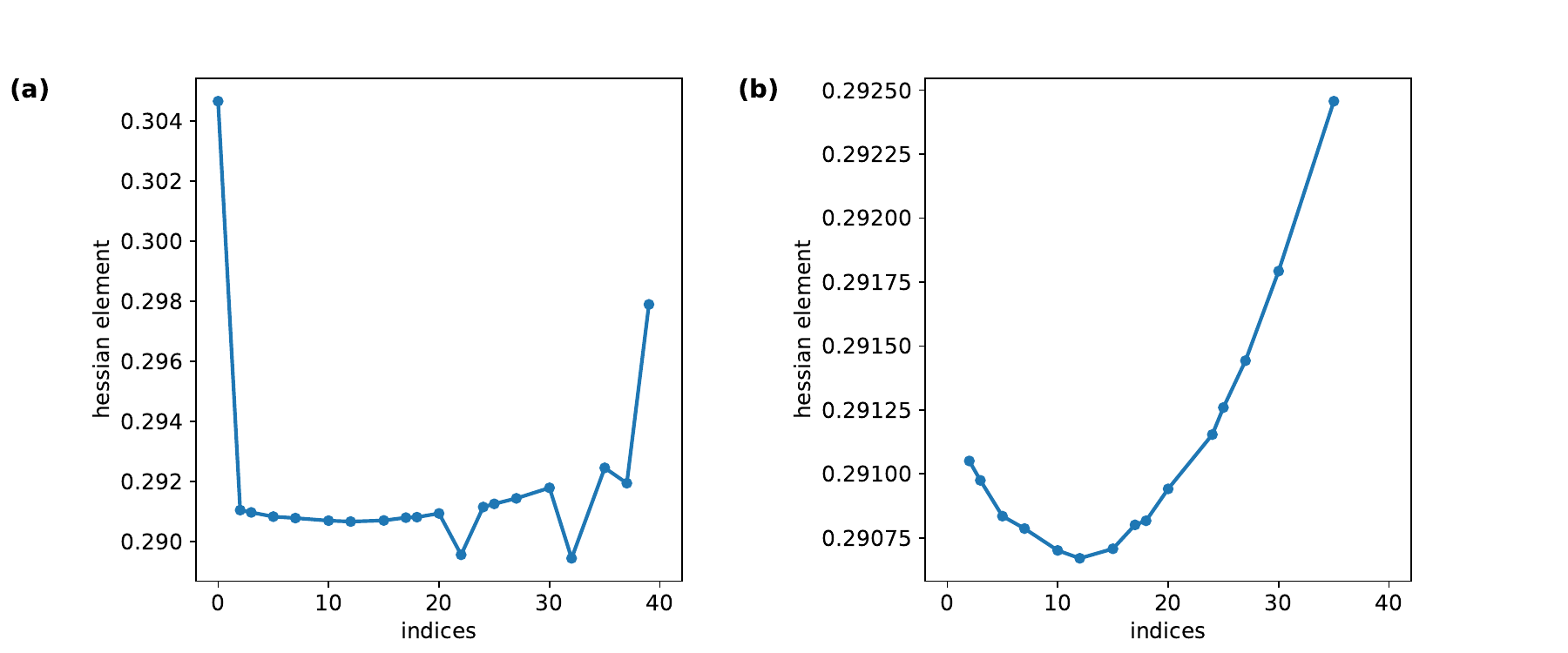}

    \caption{Illustration of the outlier removal procedure applied to Hessian matrix data. A representative Hessian matrix element along the instanton pathway of the dinitro-HBQ molecule is shown. The Hessian matrix is computed by the finite differences of the forces from density functional theory (DFT) calculations. Numerical error from the DFT calculation and the finite differencees method introduces spurious data points with large deviations, which are treated as outliers. Panel (a) shows original dataset containing outliers, whereas panel (b) shows data points after removing outliers. Applying GPR or the cubic spline method directly to the original data points in panel (a) will lead to poor model performance due to the prescence of outliers. In contrast, removing the outlier prior to modeling fitting produces accurate and reliable interpolation, demonstrating the importance of the outlier removal procedure for the subsequent machine learning analysis.
    }
    \label{fig:Outlier_removal}
\end{figure}
The Hessian matrix used for GPR is obtained from the finite differences of forces computed by DFT. The Hessian elements inevitably contain numerical noise originating from both the DFT force and finite difference approximation. In some cases, these errors produce isolated data points with exceptionally large deviations. Suce outliers can significantly degrade the performance of the regression method by biasing the fitting model toward erroneous observations. To mitigate this issue, an outlier removal procedure is applied before the model training. In figure \ref{fig:Outlier_removal}, we show one representative Hessian matrix along the instanton pathway for the dinitro-HBQ molecule. The original dataset, shown in figure \ref{fig:Outlier_removal}(a) contains several outliers arising from numerical errors, whereas these points are moved in figure \ref{fig:Outlier_removal}(b). We find directly applying GPR or cubic spline method to the original dataset results in poor interpolation quality. In contrast, training on the cleaned dataset yields smooth and physically consistent predictions, demonstrating the outlier removal is an essential preprocessing step when constructing an accurate surrogate model. 

\section{Illustration of Flexible and Rigid Molecular Modes}
\label{appendix:flexible-rigid-modes}
\begin{figure}[H]
    \centering
    \includegraphics[width=0.8\linewidth]
    {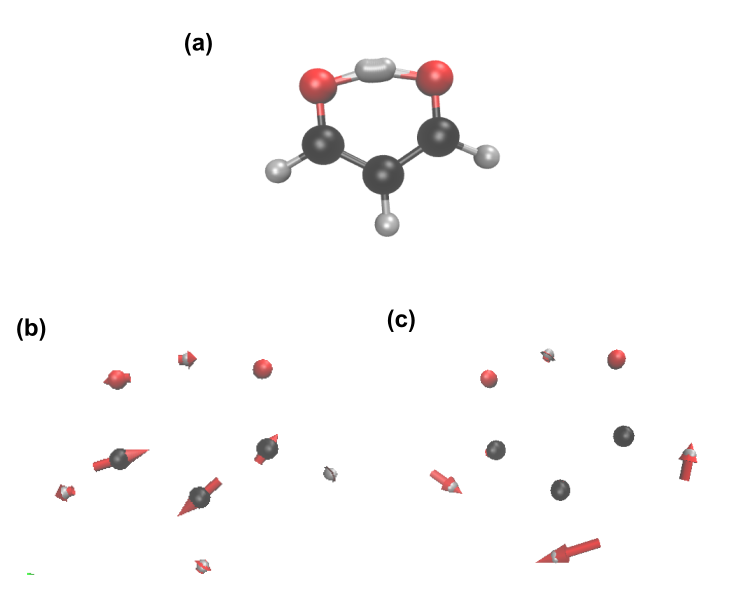}

    \caption{Illustration of flexible and rigid internal modes in the malonaldehyde for the adaptive potential energy surface regression. (a) Instanton path of the malonaldehyde molecule. (b)Representative flexible internal modes, which are strongly coupled to the proton transfer motion. The arrow length in the figure indicates the relative amplitude of atomic displacement associated with each mode. (c) Representative rigid internal modes, which are weakly coupled to the proton transfer motion. This mode is primarily associated with the motion of peripheral hydrogen atoms.  
    }
    \label{fig:rigid_flexible_modes}
\end{figure}

To illustrate the different characteristics of molecular internal modes and their implications for adaptive PES regression, we consider malonaldehyde as a representative example. As shown in Fig.\ref{fig:rigid_flexible_modes}(a), the instanton path of malonaldehyde involves a proton transfer process between two oxygen atoms, accompanied by collective atomic rearrangements of the molecular framework. The vibrational modes associated with this reaction coordinate exhibit distinct coupling strengths with the proton transfer motion.

Figure\ref{fig:rigid_flexible_modes}(b) presents representative flexible internal modes that are strongly coupled to the proton transfer process. These modes involve significant displacement of the transferring proton and therefore contribute substantially to the local variation of the PES along the reaction pathway. Accurate modeling of these degrees of freedom is essential for capturing the anharmonicity and reaction dynamics. In contrast, Fig.\ref{fig:rigid_flexible_modes}(c) illustrates representative rigid internal modes, which exhibit weak coupling to the proton transfer motion. These modes mainly correspond to the motion of peripheral hydrogen atoms and out of plane vibrations, which contribute less to the variation of the PES in the reactive region.

This distinction between flexible and rigid modes motivates the adaptive regression strategy employed in this work. By identifying the internal degrees of freedom that exhibit strong coupling to the reaction coordinate, computational resources can be preferentially allocated to accurately model the flexible subspace, while the weakly coupled rigid subspace can be treated with reduced sampling effort and the linear regression model. Such a mode-dependent strategy enables efficient construction of high-dimensional PES models while maintaining accuracy in the dynamically relevant regions.
\end{appendices}

\begin{acknowledgement}
This work was supported by the U.S. Department of Energy, Office of Science, Office of Basic Energy Sciences through the Condensed Phase and Interfacial Molecular Science (CPIMS) Program of the Division of Chemical Sciences, Geosciences, and Biosciences under FWP 80818 (C.Z.) at the Pacific Northwest National Laboratory (PNNL) and DE-SC0023249 (A.N., M.K. at the University of Washington, Seattle). A.G. acknowledges support from Computational Chemical Sciences Center “Chemistry in Solution and at Interfaces” at Princeton University funded by the U.S. Department of Energy through Award No. DE-SC0019394. N.G. acknowledges support from the U.S. Department of Energy, Office of Science, Office of Advanced Scientific Computing Research and Office of Basic Energy Sciences, Scientific Discovery through Advanced Computing (SciDAC) program under the NuQuantEm project at PNNL under FWP 85742. A.N. acknowledges postdoctoral funding from the Wenner-Gren Foundations. This work benefited from computational resources provided by EMSL, a DOE Office of Science User Facility sponsored by the Office of Biological and Environmental Research and located at PNNL, the National Energy Research Scientific Computing Center (NERSC), a U.S. Department of Energy Office of Science User Facility operated under Contract No. DE-AC02-05CH11231, and PNNL’s Institutional Computing (PIC) Program, and the Raven computational chemistry cluster at PNNL funded by the DOE under FWP 84864. PNNL is operated by Battelle Memorial Institute for the United States Department of Energy under DOE Contract No. DE-AC05-76RL1830.
\end{acknowledgement}

\bibliography{ref.bib}

\end{document}